\documentclass[letterpaper,twocolumn,10pt]{article}
\usepackage{usenix,epsfig,endnotes}
\usepackage{url}
\usepackage{color}
\usepackage{geometry,array,float,caption}
\usepackage{subcaption}
\usepackage{paralist}
\usepackage{enumitem}
\usepackage{scrextend}
\usepackage{tabularx}
\usepackage{tikz}
\usepackage{epstopdf}

\def\checkmark{\tikz\fill[scale=0.4](0,.35) -- (.25,0) -- (1,.7) -- (.25,.15) -- cycle;} 

\addtokomafont{labelinglabel}{\sffamily}

\begin{document}

\graphicspath{ {.} }

\date{}


\newcommand{\eg}{{\em e.g., }}
\newcommand{\ie}{{\em i.e., }}
\newcommand{\paragraphb}[1]{\vspace{0.03in}\noindent{\bf #1} }
\newcommand{\paragraphe}[1]{\vspace{0.03in}\noindent{\em #1} }
\newcommand{\paragraphbe}[1]{\vspace{0.03in}\noindent{\bf \em #1} }
\newcommand{\comment}[1]{\textcolor{purple}{#1}}
\newcommand{\eric}[1]{\textcolor{red}{#1}}
\newcommand{\todo}[1]{}
\renewcommand{\todo}[1]{{\color{red} TODO: {#1}}}

\title{\Large \bf FluidMem: Memory as a Service for the Datacenter}

\author{
{\rm Blake Caldwell, Youngbin Im, Sangtae Ha, Richard Han, and Eric Keller}\\
University of Colorado Boulder
} 

\maketitle



\subsection*{Abstract}


Disaggregating resources in data centers is an emerging trend. Recent work has begun to explore memory disaggregation, but suffers limitations including lack of consideration of the complexity of cloud-based deployment, including heterogeneous hardware and APIs for cloud users and operators.  In this paper, we present FluidMem, a complete system to realize disaggregated memory in the datacenter.  Going beyond simply demonstrating remote memory is possible, we create an entire Memory as a Service.  We define the requirements of Memory as a Service and build its implementation in Linux as FluidMem.  We present a performance analysis of FluidMem and demonstrate that it transparently supports remote memory for standard applications such as MongoDB and genome sequencing applications.

\section{Introduction}
\label{sec:intro}

Resource disaggregation in the datacenter is an emerging trend~\cite{costa_rethinking_2014, 1419865, Gao:2016:OSDI, han_network_2013, Lim:2009:DME:1555815.1555789, Lim:2012:SID:2192603.2192683, Ma:2014:DVM}. In particular, memory disaggregation is a focus of much of the recent work, in which remote memory is made available to improve the performance of applications in the datacenter.  As an example, scale-up applications are those that require more resources on a single node to process larger problem sizes. These applications may be ill-suited for scaling-out in a distributed fashion either because the computation is serial in nature, or the code base is viewed as legacy, where the effort or expertise required for parallelization is infeasible. 
Memory intensive scale-up applications for scientific~\cite{Luo2012,zerbino:velvet:2008} and business analytics purposes~\cite{Farber:2012:SHD} reach the upper limit of memory in a single system. In contrast, memory disaggregation in a data center rack-scale architecture~\cite{Gao:2016:OSDI,han_network_2013,Lim:2009:DME:1555815.1555789,Lim:2012:SID:2192603.2192683} would free memory intensive applications to make allocations of memory capacity from other nodes connected by high-speed networks.

New technologies are making it feasible to achieve memory disaggregation.  We have seen a convergence of advances in data center network transports like HULL~\cite{alizadeh2012hull}, a new interface to the Linux kernel paging mechanism~\cite{www:userfaultfd}, and the development of high-speed key-value stores like RAMCloud~\cite{ousterhout:transactions:2015}.  Together, these technology trends provide a path forward to offer \emph{memory as a service} in today's datacenters.

\begin{figure}[t]
\vspace{-0.1in}
	\begin{center}
		\includegraphics[width=0.5\textwidth]{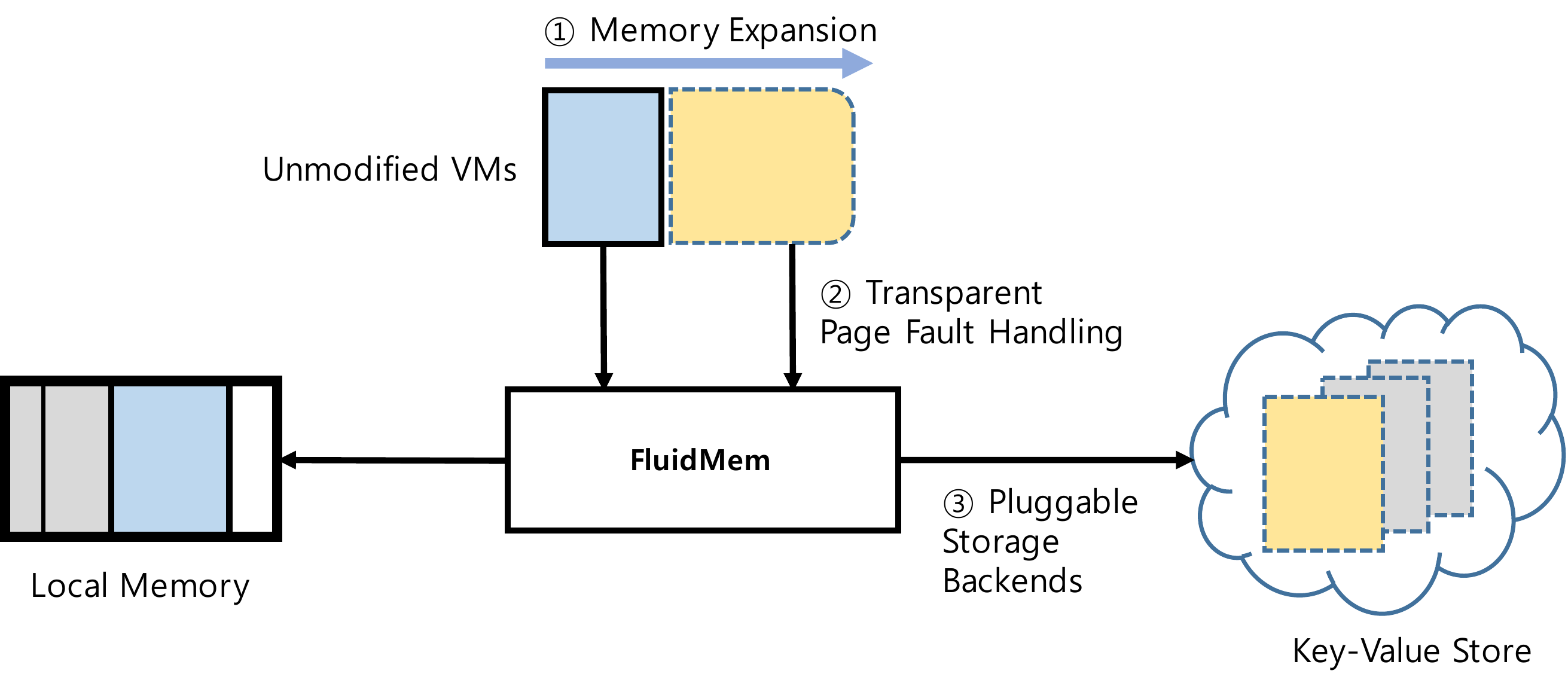}
	\end{center}
	\vspace{-0.2in}
	\caption[FluidMem Architecture]{
	FluidMem Architecture
}
	\label{fig:fluidmem_arch}
	\vspace{-0.2in}
\end{figure}

The latest work in memory disaggregation~\cite{Gao:2016:OSDI} demonstrated the feasibility of remote memory by implementing a prototype using a swap device to emulate remote memory, and also evaluated the performance of Spark~\cite{zaharia2012spark} and COST~\cite{McSherry:2015:SBC:2831090.2831104} in a disaggregated datacenter environment. Their conclusion on the network requirements for datacenter disaggregation, was that network latencies of 3-5$\mu$s between a VM and disaggregated memory stores were necessary for minimal performance degradation.  Their system assumptions included partial CPU-memory disaggregation, a cache coherence domain that is limited to a single compute blade, page-level remote memory access, and a VM abstraction for logical resource aggregation. That work however falls short as it is basically a simulation rather than a complete system, relies on specialized network technologies, and does not consider larger issues for deployment in the cloud, such as support for multiple back-end memory stores as well as support for cloud operators in addition to cloud users.

\begin{figure}[t]
\vspace{-0.1in}
	\begin{center}
		\includegraphics[width=0.5\textwidth]{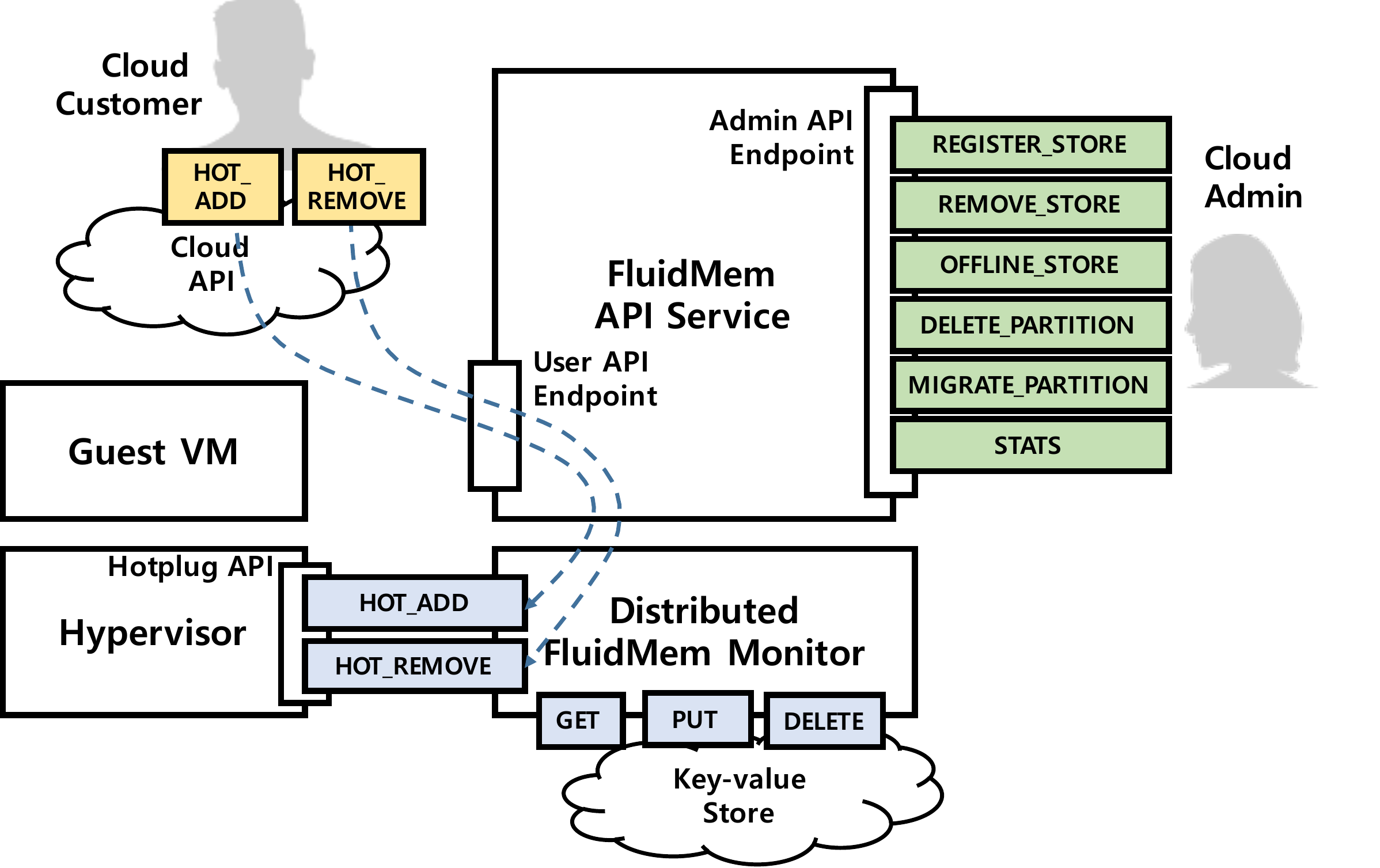}
	\end{center}
	\vspace{-0.2in}
	\caption[Memory as a Service Architecture]{
	Memory as a Service Architecture
}
	\label{maas_arch}
	\vspace{-0.2in}
\end{figure}

In this paper, we present FluidMem, a complete system to realize a
disaggregated memory in the datacenter. Its architecture is show in
Figure~\ref{fig:fluidmem_arch}. Going beyond simply demonstrating remote memory is feasible, we create an entire Memory as a Service.  What this means is that both scale-up and scale-out applications and VMs can transparently access remote memory on demand via a working implementation that supports current operating systems used in cloud datacenters, as well as today's heterogeneous datacenter networking and storage technologies.  To achieve flexibility in supporting heterogeneous backends and networks, we adopt the \emph{userfaultfd} mechanism in Linux.  To allow VMs to expand memory dynamically, we leverage the \emph{hotplug} mechanism in Linux.  The complete system is shown in Figure~\ref{maas_arch}.

We have designed and implemented FluidMem to enable transparently expanding memory capacity of cloud VMs, providing Memory as a Service.  Our paper makes the following contributions:

\begin{enumerate}[noitemsep]
	\item{} Defines the memory as a service abstraction, and architects a system to implement it for use in today's datacenters.
	\item Demonstrates the ability to transparently provide remote memory to standard applications such as MongoDB, two genome sequencing applications, the Graph500 benchmark suite of applications, and Spark.
	\item{} Demonstrates the feasibility of running applications inside a VM with more physical memory than what can be provided by a single node.
	\item Exposes remote physical memory to the application (through hotplug), enabling those applications either to avoid failure due to out of memory conditions or perform appropriate optimizations using the extra memory capacity.  
	\item{} Demonstrates the flexibility of user-space page fault handling by supporting three different key-value stores for backing memory storage and allowing optimization such as concurrent and asynchronous page fault handling.
\end{enumerate}

In the following we discuss the challenges of realizing Memory as a Service, define its requirements, then present the FluidMem implementation of Memory as a Service, followed by an evaluation of FluidMem.


\section{Challenges for Memory as a Service}
\label{sec:challenges}

\subsection{Inserting a remote memory abstraction}

Our first challenge is to determine at what point in the handling of a memory request do we insert a remote memory abstraction.
In virtualized systems there are many steps between a load or store instruction originating from the application and ultimately accessing a physical medium. If the memory area allocated by the application is a normal anonymous region, then there will be a page fault within the guest, where the virtual address is translated into a guest physical address. If the page is resident and not on the swap device, the hypervisor will translate the guest physical address to a host physical address. Note for fully virtualized systems, sometimes this will involve an intermediate step of guest physical address to host virtual address translation. Modern para-virtualization hypervisors such as Xen and KVM support CPU functionality that performs this intermediate step. 

For a cloud computing environment targeting the support of a wide range of
applications and the ability for users to bring their own operating systems,
we believe it is essential to not require either the guest VM or the
application itself to be modified to leverage remote memory. This rules out
solutions used in high-performance computing such as
PGAS~\cite{Chamberlain:2007:PPC,Charles:2005:XOA,nelson_latency-tolerant_2015} and direct usage of key-value stores~\cite{dragojevic_farm:_2014,ousterhout:transactions:2015}.
Remote memory systems that interpose standard libraries~\cite{midorikawa:cluster:2008} for virtual memory accesses are also unacceptable because the necessary toolchain for compiling and linking may not be available. Using a remote memory system that modifies the guest VM kernel to handle page faults differently for local versus remote accesses is also not a useful solution for the cloud. The implementation would be tied to a specific kernel, and only one operating system, thus severely limiting use cases in the cloud.

A reasonable alternative would be leveraging the swap interface of the guest kernel. The necessary modifications to enable remote memory would be contained in a kernel-space driver loaded as a module on the hypervisor.  The guest kernel is able to access this driver through the standard block interface when swapping in or out a page. However, a drawback of this approach is that some pages cannot be swapped out, such as those belonging to the kernel, explicitly pinned pages, or huge pages in Linux. Additionally, every VM would instantiate a device using this driver, and each needs to be back-ended by some sort of remote memory mechanism. It is not clear how the management of so many devices would scale in a cloud environment with thousands of VM’s.

Another option at the lowest level is to modify the hypervisor’s kernel to
handle page faults to remote memory. Such an approach was used with kernel
implementations of DSM~\cite{fleisch:spe:1994,souto:atc:1997}. A difficulty
with this however is that the page fault handling code in Linux and other
operating systems is core to the kernel, and cannot be separated into a
module like the code for a driver backing a swap device. We would not expect
cloud providers to willingly run a kernel maintained downstream,that will
lag behind in releasing security fixes. Furthermore, the complexity and
effort to implement Memory as a Service in kernel-space would be immense.
While a modified kernel could trap to user-space, as with
\emph{userfaultfd}~\cite{www:userfaultfd}, it is unlikely that the Linux
community would accept a second mechanism when \emph{userfaultfd} already
exists. Previous kernel-space attempts at remote memory~\cite{Chapman:2009:VVS,fleisch:spe:1994,www:kerrighed,souto:atc:1997} have failed to gain acceptance as part of the Linux or FreeBSD kernels, and as a result, have not maintained compatibility with current versions.

We believe a better solution is to trap accesses within the user-space memory regions on the hypervisor. Since the Linux 4.3 kernel, an interface called \emph{userfaultfd}~\cite{www:userfaultfd} is available to notify user-space of page the faults in designated virtual address ranges. 
Additionally, \emph{userfaultfd} supports shared memory, \emph{tmpfs}, and huge page memory regions. An extension of the \emph{ioctl} system call is used to handle moving pages in and out of kernel-space, and waking up the process block on the page fault. Using this mechanism, a single process on the hypervisor can monitor a list of \emph{userfaultfd} file descriptors while making use of remote memory storage. 
Since cloud providers can choose to standardize on the hypervisor used across a cloud infrastructure, we believe minimal modifications at this level are acceptable. While this implementation choice is limited to Linux hypervisors, this includes the majority of hypervisors running in clouds today. 

\subsection{Designing flexibility}
Every cloud data center cannot be expected to have homogenous systems or network infrastructures. Nor can it be expected that any two cloud providers share similar deployments. There could be differences in the physical network technology, for example public cloud providers are most likely to rely on Ethernet, but private-clouds for scientific purposes may be willing to deploy an Infiniband fabric.
Furthermore, alternative network transports the may be chosen for Ethernet, such as the user-space framework DPDK. In these cases, it is desirable for the page fault handler to also be in user-space since DPDK for Ethernet and ~\emph{libibverbs} for Infiniband make use of kernel-bypass to avoid extra context switches. Since network latency is a significant factor in overall page fault time for remote memory, supporting these high-speed frameworks is a motivation for us to design a user-space page fault handler.

Another challenge is making the backend storage for remote memory capable of satisfying the variety of service requirements that cloud providers have. Since the paging of many applications may be handled by a single framework, it must handle concurrent operations, support low latency reads and writes, scale in capacity and throughput, and provide durable storage. Satisfying all of these requirements in a specialized distributed memory store built into the kernel would be an enormous undertaking. It would be prone to bugs, have security implications for the entire operating system, and be unlikely to be accepted by maintainers of open source kernels such as in Linux.  We design our page fault handler to flexibly support a variety of backend stores
in user-space.
Distributed in-memory key-value stores such as RAMCloud~\cite{ousterhout:transactions:2015} and FaRM~\cite{dragojevic_farm:_2014} are complex distributed systems in their own right, but since they have focused on the intended use case of high-performance in-memory storage in user-space, they are more feature complete than the memory stores of previous works. As an example, RAMCloud provides crash-recovery, tolerating node failures without loss of availability to the data store. 

\subsection{Two user groups to please}

To achieve widespread adoption of Memory as a Service in cloud computing environments, we must accommodate the needs of both cloud users and cloud operators as well as the trade-offs they are willing to tolerate.
On the user side, the most visible qualities will be performance, ability to elastically add memory, ease-of-use, and service reliability. The cloud user may be willing to tradeoff performance for improved durability of the remote memory store for example.  In this case,
we could leverage a backend store such as RAMCloud, which offers durability of pages through replication.
Users of cloud services such as Amazon's Elastic Block Store (EBS) have already demonstrated their tolerance of suboptimal performance for flexibility benefits such as attaching the storage device to any cloud instance, in dynamic capacity provisioning.
On the other hand, cloud operators will be concerned about implementation and management complexity weighed against any efficiency gains from aggregation and the service improvement that can be provided to a customer. Some of the prior memory disaggregation work~\cite{han_network_2013,Gao:2016:OSDI} discusses that there are utilization efficiencies to be had benefiting the operator, where VMs don't need to be over provisioned for worst-case scenarios. Therefore, Memory as a Service should provide mechanisms to ease the provisioning and deployment of VMs when dynamic allocation of remote memory is desired.

\section{Defining the requirements for Memory as a Service}


In order to bring benefit to cloud computing environments, Memory as a Service must be more flexible than previous remote memory systems to provide a generalized service to users while being compatible with heterogeneous cloud infrastructures. We have distilled the service needs down to five requirements that define Memory as a Service. 


\begin{enumerate}[noitemsep]
	\item{} No code modifications to applications or the methods to compile them. The interface to remote memory must adhere to the same load and store interface of local memory
	\item{} Current operating systems used in cloud data centers should be supported
	\item{} Heterogeneous networks and memory storage systems should be supported.
	\item{} Each cloud user should be unaware of other users that may be using the memory service
	\item{} It is feasible for cloud operators to manage the extra complexity. There must be some economies of scale as the number of users of the service increases.
\end{enumerate}

\subsection{Unmodified Applications}
Cloud computing does not impose demands that applications must be modified to be moved from one cloud to another. In fact, one of its strengths is that applications can run anywhere. Thus, in keeping with the spirit of cloud computing, applications must be able to benefit from Memory as a Service in their unmodified form. The use of remote memory as opposed to local memory must be transparent to the application, with the underlying Memory as a Service infrastructure handling the complexities of using a remote memory. This means that memory-intensive applications ranging from legacy to proprietary to cutting edge will benefit from Memory as a Service.


\subsection{Current Operating Systems}
The need to expand memory capacity dynamically is a need of today's applications~\cite{Farber:2012:SHD,Luo2012,zerbino:velvet:2008} so Memory as a Service should be an offering of today's cloud platforms running these applications. Both Linux and Windows have established footholds as the hypervisor's in public clouds without significant change in over 10 years. So, the practicality of Memory as a Service depends on its compatibility with current operating systems.

Related to the choice of operating systems used in cloud data centers, the hypervisor used is unlikely to change. Cloud operators have been building engineering expertise and service offerings built on their chosen hypervisor for over 10 years in some cases. It would be unreasonable to expect them to change to a newly developed hypervisor that hasn't been hardened in production like KVM~\cite{www:kvm} or Hyper-V~\cite{www:hyper-v}.

A separate issue though, is the choice of operating system for the guest VM running in a cloud. Since hypervisors decouple this choice from the host operating system, cloud users are free to choose from a multitude of different machine images with popular operating systems (for a given micro-architecture). Memory as a Service should preserve this flexibility, and allow different operating systems, both Linux and Windows to use the remote memory it provides. This means that remote memory must be transparent to the OS kernel, for modification of every kernel in use by cloud customers would be infeasible.

\subsection{Heterogeneous Hardware}
Memory as a Service should not prescribe a particular network infrastructure or backend memory storage systems, but should instead support heterogeneous mix present in today's datacenters. This means that Ethernet and the conventional TCP transport should be supported for moving pages in and out of remote memory, but other transports should be supported as well. For example, those that support RDMA and/or kernel-bypass.

Additionally, there should be flexibility in the choice of a backend memory storage system. This thesis builds the case for a distributed memory storage system in software, but the specific system can vary with cloud provider's preferences or by where engineering expertise has accumulated for specific key-value stores. A cloud provider may opt to even have multiple memory backends supporting different guarantees for memory reliability and performance. Memory as a Service should be agnostic to the specific type of storage system used, and should interface with generic and simple operations.

\subsection{Service Isolation}
A public cloud is almost certain to be a multi-tenant environment, where multiple users can be assigned VMs that are co-located on the same physical server. Users should feel sufficiently isolated from each other, such that the activity of one customer's VM does not adversely affect the performance of another VM. With respect to Memory as a Service, this means that the hypervisor should be able to handle heavy paging activity from one VM, with sufficient overhead should other VMs on the same hypervisor begin intensive paging as well. The acceptable oversubscription ratio of maximum memory traffic from all VMs to the remote memory paging capacity of the hypervisor will be specific to a cloud and the guarantees it provides customers. However, all Memory as a Service implementations should be able to scale aggregate handling capacity per hypervisor with the number of VMs running. Additionally the backend must be capable of handling memory traffic from many VMs concurrently. The scaling can be capped to avoid excessive degradation, but it must certainly be greater than a single VM per hypervisor.

Similar to performance isolation, Memory as a Service should also provide isolation for security purposes. A malicious user of one VM should not be able to snoop or redirect the remote memory traffic of another VM. Additionally, listings of resources shown to one tenant should not divulge information about other tenants using the memory service.

\subsection{Managed Complexity}
The number of VMs managed in a single day center can number in the thousands or tens of thousands. Operating Memory as a Service cannot mean that 10,000 SSD drives need to be installed, mapped to VMs, monitored, and replaced upon failure, or the cloud provider would never offer such a service. A system that aggregates memory storage as a distributed system that can be managed as a logical entity, and tolerate individual hardware failures without affecting availability, would be much better. Practical memory as a service implementations must provide â``administrative economies of scale''.

\section{FluidMem Implementation}
\label{sec:implementation}
\begin{figure}
  \includegraphics[width=\linewidth]{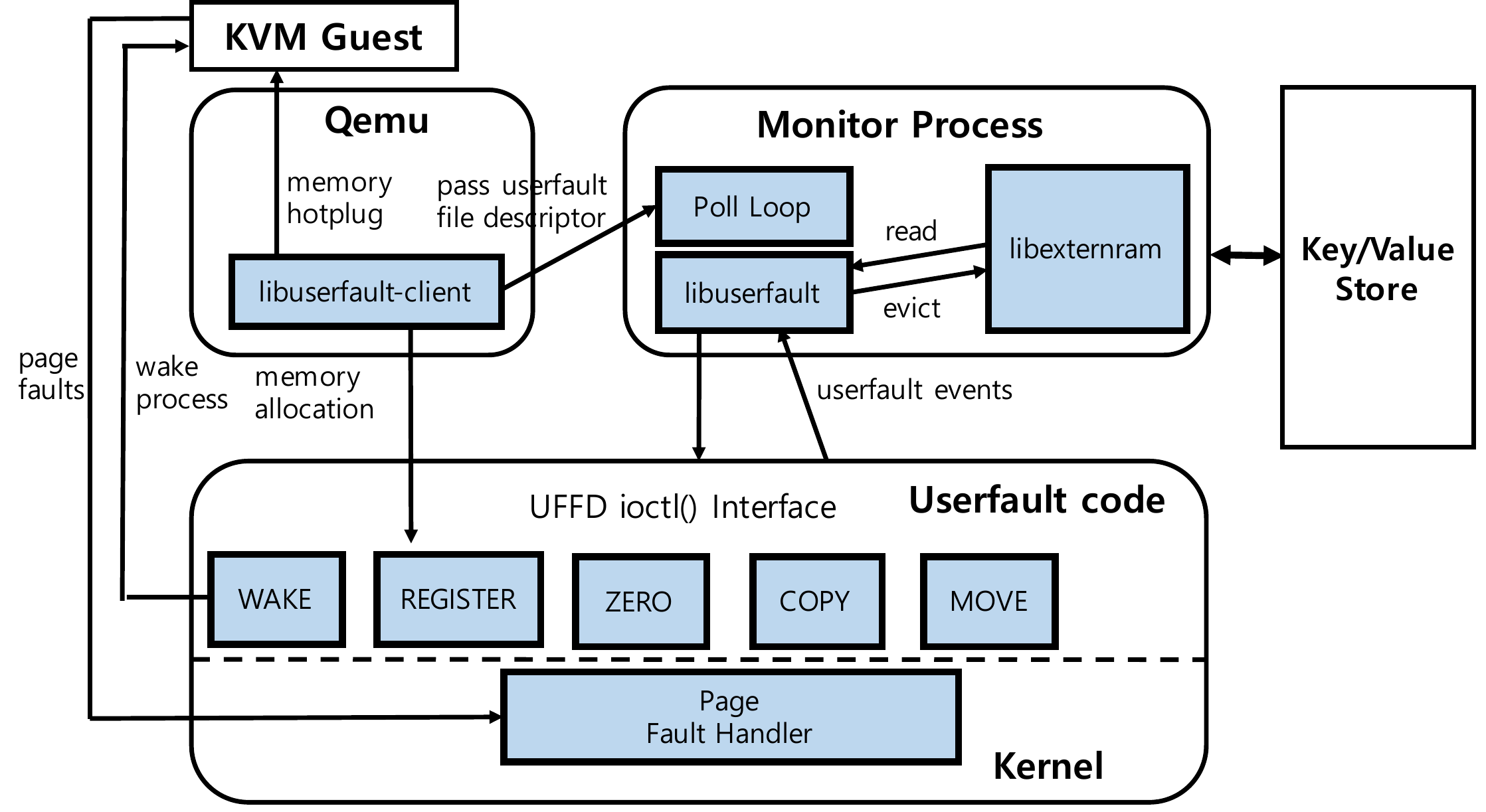}
  \caption{FluidMem Implementation}
  \label{fig:fluidmem_impl}
\end{figure}

There are three main elements that work together in our implementation of Memory as a Service, called FluidMem.  First, there are the modifications to the hypervisor that make use of the hotplug interface to pass in memory regions registered with the \emph{userfaultfd} kernel interface. Second, we modified and extended the \emph{userfaultfd} interface to move pages out of the guest VM when evicting a page to remote memory. Third, there is a daemon on each hypervisor called the monitor process to handle trapped page faults and service them from remote memory as needed. It maps pages from different applications to a partitioned key/value store.

\subsection{Hotplug modifications}

Memory hotplug was originally developed for high reliability servers to add and remove DIMMs with a live operating system.  The kernel manages mapping of page frames to the newly added physical address space.  In virtualized environments, this functionality can be used to add memory to a VM from outside its original memory allocation.  

Since \emph{qemu} (the user-space component of the KVM hypervisor) already supports memory hotplug, my modifications were limited to patching \emph{qemu} to register the memory allocation with \emph{userfaultfd} and pass a file descriptor to the monitor process, as shown in Figure~\ref{fig:fluidmem_impl}. The methods to accomplish this are encapsulated in a dynamically linked library that called \emph{libuserfault-client} that can be linked with \emph{qemu} or any other user level application.

By adding the new file descriptor to a list that it continually polls, the monitor process will be notified of page faults within this new memory region. A challenge with this mechanism is that the file descriptor associated with the \emph{userfaultfd} region must move from \emph{qemu}'s address space to the monitor's address space. This is accomplished with \emph{libuserfault-client} by setting up a connection between \emph{qemu} and the monitor by using a Unix domain socket.  

\subsection{Userfault handling}

When a new VM starts up with a hotplugged memory region, the monitor process becomes aware of a new file descriptor to poll by the process described in the previous section.  Later, when a guest VM writes to a page in that region for the first time, the page fault handler in the Linux kernel passes off execution to the ~\emph{userfaultfd} code, which creates an event on the file descriptor and blocks the guest VM. Monitor's polling is interrupted by this event, and it reads the address of the page fault from the file descriptor. Since this is a new page, the monitor returns a copy-on-write zero page to the VM with the ZERO \emph{ioctl}.  

Eventually, the page that was given back to the VM will need to be evicted to a key-value store to make room for pages that were more recently faulted in. In order to support this operation, we extended the \emph{userfaultfd} kernel interface to signal that a page should be removed from the VMs address space.  This leaves a ``hole" where subsequent accesses will trigger another page fault. This frees the monitor to transfer the contents to an external key-value store.  Consequently, no intervention is needed from the guest, and the monitor process can issue the MOVE \emph{ioctl} when memory in the hypervisor needs to be freed.


When a guest re-accesses a page, FluidMem employs the COPY \emph{ioctl} in the other direction to retrieve pages from an external key/value store and insert them into the guest.



\subsection{The monitor process}
The monitor's primary responsibility is to watch for page faults by monitoring a list of file descriptors and responding appropriately.  Each page fault event includes information on the address and whether the fault is a read or write.  
The monitor process handles each event in a poll loop, deciding whether to return a zero page or fetch the page from external storage (externram).  Reads of pages that still remain in the VM do not trigger a page fault.  In the case of handling a read page fault where there is deemed no more space left for the page pulled from external key-value store, the monitor process will choose a page to evict and invoke the \emph{userfaultfd} MOVE mechanism to transfer the page from the guest VMs hotplug memory to the key-value store.

A write fault is always served with a zero page.  Even if the page exists in external key-value store, a new copy is created on the hypervisor.  Since the VM will only ever access this page through the monitor, it is guaranteed to have the most up-to-date copy.  On eviction, the newly written page will replace the old page in the external key-value store.

The monitor's code organization is structured such that it has an API to interact with an external key-value store.  The currently supported operations are read, write and remove.  These are packaged in a \emph{libexternram} library that is then called by the \emph{libuserfault} library.  This structure of linked libraries allows for different key-value store implementations to be written in various languages.
FluidMem supports three key-value stores in \emph{libexternram}: RAMCloud, memcached, and an in-memory hash structure local to the monitor.

There are other events the monitor is responsible for that are outside of the poll loop.  Additional threads are spawned for registering new userfault regions or detecting zombie processes whose file descriptors should be removed from the list.

\subsection{The key-value store}
We focus our implementation primarily on the version of FluidMem that incorporates RAMCloud as the key-value store backend.  The reasons for this are that RAMCloud stores all primary copies of data in DRAM for fast access, supports replication without the need for waiting for a write completion, the partitioned key space is hashed over multiple nodes so our storage capacity grows with the number of servers that are part of the RAMCloud cluster, and performance scales since partitions are hashed over the nodes in the cluster.  Another benefit is that RamCloud uses high speed network transports with kernel bypass, e.g. Infiniband and DPDK over Ethernet. 
We also integrated two other key-value stores as part of FluidMem, namely memcached~\cite{www:memcached}, which uses the kernel's TCP/IP network stack, and a C++ map structure.  We did this to demonstrate the flexibility of our user-space approach.

\subsection{Implementation optimizations}
\label{sec:implementation_optimization}

The monitor process implements a variety of performance optimizations.  First, since we're dealing with remote memory that has a higher latency penalty than local memory, many existing caching strategies apply.  We maintain a structure containing metadata on the pages that are part of the hotplugged memory region, but reside in the guest VM's address space.  The structure is dual-indexed as both a hash list and an LRU list for efficient lookups and identification of pages to evict.  The size of this structure can be adjusted in response to memory demands on the hypervisor. 

In order to exploit the benefits of locality of page accesses in many applications, we adopt a cache for pages. The page cache stores the pages in memory and delivers the pages to the guest VMs when requested with much less latency than the remote memory. Which pages to store in the page cache is decided by our prefetch mechanism. We retrieve redundant pages that are expected to be used in the near future in addition to the original requested page when we retrieve a page from key/value store. However, the hit ratio was not so high in our tested general applications, so we use a conservative prefetching mechanism: prefetch a single page if adjacent pages are requested consecutively. This guarantees that we prefetch a page only when the program accesses the pages in a sequential way. 

The evicted pages do not need to written to the key/value store immediately as long as they are requested by the guest VM. Therefore, we store the pages to evict in a list and write to the key/value store in a batch using a separate thread rather than immediately writing every single page. In this asynchronous evict mechanism, we utilize the benefits of multi-write operation which is faster than multiple write operations with the same number of values in many key-value stores. For prefetching, we also use a separate thread and multi-read operation in order to exploit the parallelism between threads and short latency. 

The monitor process checks whether an evicted page is an all-zero page before writing to key/value store, thus avoiding unnecessary network transmissions. If it is an all-zero page, it marks in the data structure and does not write to key/value store. If the page is requested next time, the monitor process simply returns a new all-zero page.

We also found that there are frequent context switches when assigning a temporary memory buffer, i.e., \emph{mmap} call, in handing a page fault in monitor process. To avoid this inefficiency, we assign a separate thread that re-initializes the array of temporary page buffers before they are used by page fault handing functions. This re-initialization thread operates for the temporary page buffers used by page evict and read handling functions.

Finally, we observed that certain threads of monitor process such as asynchronous evict/prefetch threads show a high CPU utilization than other threads. In order to avoid that the performance is bottlenecked by single CPU capacity, we run these threads on different CPU cores than other threads by setting the CPU affinity for each thread. We test the performance benefits of different performance optimization mechanisms mentioned above in Section~\ref{sec:microbenchmarks}.

\subsection{Hotplugging in a cloud environment}
Cloud environments handle VM life-cycle events such as launching, terminating and resizing through API calls. To accommodate cloud operators, FluidMem's memory expansion capabilities are integrated into the OpenStack~\cite{www:openstack} cloud framework. Users (end-users or operators) interact via a webpage or CLI to instantiate VMs with OpenStack's \emph{nova} service, which was modified  to instantiate hotplug-enabled VMs. The \emph{nova} service interacts with the hypervisor through the \emph{libvirt} API, which has been modified to specify whether a \emph{qemu} VM should be started with hotplug memory or not. If hotplug memory was specified, then \emph{qemu} will register the region with \emph{userfaultfd} and pass the region's file descriptor to the monitor process using \emph{libuserfault-client}.

If additional remote memory is desired after boot, a Python script carries out the necessary \emph{libvirt} API calls, which in turn call \emph{qemu}'s memory hotplug functions. While from the guest's perspective memory hotplug is typically divided into 2 steps, adding the memory, and then on-lining it, many default Linux distributions include \emph{udev} rules that automatically online the memory once it becomes available. In this way, the guest will see the hotplug memory as part of available memory when running a command such as \emph{top} or \emph{free}.







\section{Evaluation}
\label{sec:eval}

In this section, we analyze the latency of FluidMem components.  Next we consider the performance of different optimizations of FluidMem.  Finally, we describe use cases of standard applications running transparently on FluidMem, demonstrating that exposing remote physical memory to an application enables it to improve its optimizations (MongoDB) or avoid failure by having sufficient memory to run to completion (genome applications).


\subsection{Performance analysis of FluidMem components}

\begin{figure}
	\includegraphics[width=\linewidth,height=8cm,keepaspectratio]{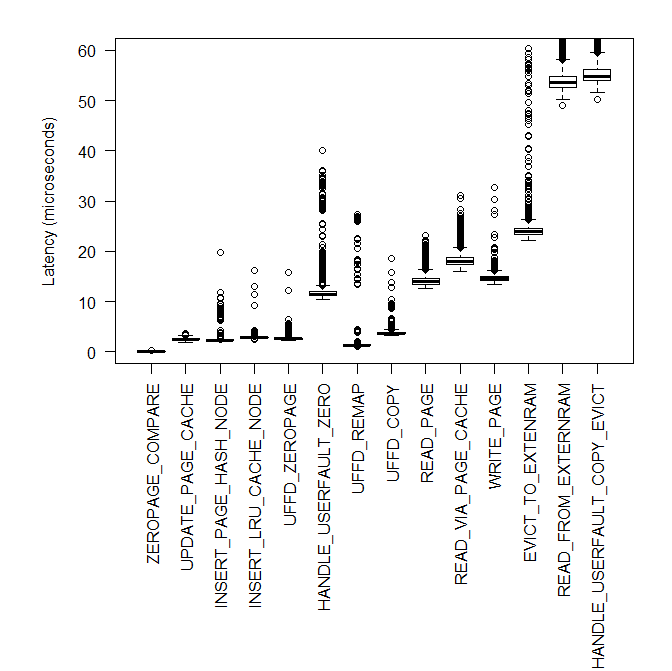}
	\centering
	\caption{Distribution of latencies for individual components of FluidMem.}
	\label{fig:boxplot}
\end{figure}

Shown in Figure~\ref{fig:boxplot} is the distribution of latencies for various scopes of FluidMem code paths. We instrumented the FluidMem monitor process to time the contribution to overall page fault latency for noteworthy sections of the code. For timing purposes, we removed the asynchronous prefetch and eviction code. This means that most of the code sections are inclusive of the code sections to its right. The exception is that to code paths diverge depending on if a page is seen for the first time (a zero page is returned), or if it has been seen before (causing a write to RAMCloud on eviction, a read from RAMCloud, then copying the page from user-space to kernel-space). The overall page fault latency for the former case is represented by \emph{HANDLE\_USERFAULT\_ZERO}, which has a median just over 10 microseconds, because it does not involve a RAMCloud operation. The latter case, however, involves both a read and write to RAMCloud and is represented by the rightmost boxplot~\emph{HANDLE\_USERFAULT\_COPY\_EVICT}. Included in the timing for the rightmost boxplot is the function \emph{READ\_FROM\_EXTERNRAM}, which in turn includes \emph{READ\_VIA\_PAGE\_CACHE} and \emph{EVICT\_TO\_EXTERNRAM}.  Those two components are mutually exclusive, where \emph{WRITE\_PAGE} and \emph{READ\_PAGE} respectively are the largest contributors to delay.  The three core \emph{userfaultfd} methods are \emph{UFFD\_ZEROPAGE}, \emph{UFFD\_COPY}, and \emph{UFFD\_REMAP}.  These are issued through an \emph{ioctl} system call and will incur a context switch into kernel-space.  These are the main unavoidable delays in FluidMem's design, but as can be seen in the Figure, they are insignificant compared to the delay of reading from an external key value store.

Some functions shown in the boxplot on the left are related to our cache design, but only demonstrate modest contributions to latency. This indicates that further optimizations in cache be design will not significantly improve overall performance. Since reading and writing to RAMCloud or the largest contributors to latency, optimizations to code that batch pages together and exploit options for concurrency should be continued.

Of note is the leftmost boxplot, which is time around the code to compare whether a page contains all zeros.  If the page is found empty, an optimization was implemented to avoid writing the page to RAMCloud. Whether this optimization is beneficial to a workload depends on the percentage of pages that are evicted containing all zeros. A ratio less than 1/100 would be beneficial. While optimization such as this may not be beneficial in the general case, they are easily implemented in user-space, and can be turned on for particular workloads.






\subsection{Latency micro-benchmarking}
\label{sec:microbenchmarks}

To deeply understand the effects of different performance optimization mechanisms presented in Section~\ref{sec:implementation_optimization}, we ran micro-benchmarks that timed between the entry and exit points in the kernel's page fault handler while running a simple test program that reads from and writes to a memory region. The test program is linked with our \emph{libuserfault} library, so a memory region is registered with the monitor process without the involvement of the virtualization layer. The test program repeatedly accesses an entry of an array with the size of assigned memory region and if it is already initialized, it prints the value to the screen, increases the value, and if not, it initializes the entry with a random value. The sequential application chooses the entry to access sequentially, while the random application selects the entry randomly. We also implemented a function to measure the times consumed in several critical function calls such as page fault handling. 

Table~\ref{tab:microbenchmark_result} shows the average latency of a userfault handling for different performance optimization mechanisms in sequential and random test applications. We applied different mechanisms in a cumulative way: for example, +prefetch represents the case when page cache, zero page optimization, and prefetch are applied. As shown in Table~\ref{tab:microbenchmark_result}, when all the optimization mechanisms are applied, the average latency is significantly reduced (by 53.7\% and 30.7\% for sequential and random applications, respectively). Several optimization mechanisms such as page cache and prefetch increase the average latency depending on the application due to their overhead. However, these mechanisms are required for other mechanisms such as zero page optimization, asynchronous prefetch, and when these additional mechanisms are combined, the overall performance is greatly improved. Some mechanisms such as zero page optimization and CPU affinity offer a significant performance enhancement in one application but a small performance degradation in another application. We include these mechanisms since the overall performance can be improved with these mechanisms for general applications whose access patterns are a combination of random and sequential accesses.

\begin{table}\scriptsize
	\renewcommand{\arraystretch}{1.8}
	\caption{Average latency of a userfault handling for different optimization algorithms in sequential and random applications (units: microseconds).} \label{tab:microbenchmark_result}
	\vspace{-0.05in}
	\centering
	\begin{tabular}{|c||c|c|}
		\hline & Sequential & Random \tabularnewline \hline \hline
		Default & 75.535 & 69.785 \tabularnewline \hline
		+Page cache & 72.37 & 78.4 \tabularnewline \hline		
		+Zero page optimization & 55.245 & 78.835 \tabularnewline \hline		
		+Prefetch & 59.73 & 86.205 \tabularnewline \hline		
		+Asynchronous eviction & 36.855 & 66.45 \tabularnewline \hline		
		+Asynchronous prefetch & 36.055 & 63.765 \tabularnewline \hline		
		+CPU affinity & 39.675 & 50.595 \tabularnewline \hline		
		+Asynchronous re-initialization & 34.94 & 48.325 \tabularnewline \hline		
	\end{tabular}
\end{table}

\subsection{Application use cases that leverage FluidMem}

This section describes use cases of standard applications running transparently on FluidMem, demonstrating that exposing remote physical memory to an application enable better optimizations (MongoDB) prevent failure from insufficient memory (genome applications).

Applications that will benefit the most from FluidMem are those that load large datasets or indexes into memory. With FluidMem the amount of physical DRAM on a server is no longer the upper bound for the amount of addressable physical memory, so the application's working set size is free to grow beyond local DRAM and spill onto remote memory. Some applications can easily partition their working set into discrete chunks that will fit into the RAM available, and as such, are not likely to see benefit from adding more memory. However, if the partitioning method is not known before hand, then FluidMem can improve performance by allowing more of the dataset to be resident in RAM. While partitioning of the data for distributed operation may be possible with explicit refactoring of the application's code, FluidMem provides an alternative solution to load these datasets into memory with requiring development effort from system engineers.

\subsubsection{MongoDB}
To quantitatively demonstrate the basic benefit of storing larger working sets in cluster memory using FluidMem, this evaluation looks at MongoDB as an example of an in-memory application that uses RAM as a caching layer. 
MongoDB is a document store commonly used for cloud applications that facilitates fast retrieval of data stored on disk by caching its working set of documents in-memory. There are two storage engine options available for persistence, one that uses \emph{mmap} to let the kernel manage which pages are cached in memory, and a newer engine ``WiredTiger'' that uses a combination of an application-specific cache and the kernel's filesystem cache. This evaluation is done with the \emph{mmap} storage engine because its simple cache design and the broad applicability of the \emph{mmap} strategy for caching data in memory. 
The evaluated MongoDB configuration consisted of a single server, rather than a sharded cluster.

The workload chosen for evaluating MongoDB with FluidMem was derived from the Yahoo Cloud Serving Benchmark (YCSB) suite~\cite{Cooper:2010}, and consisted of scan requests of 10 records, where each record is 10 KB.  The workload was run from a separate server issuing sequential scan requests to the MongoDB VM. In this way, data is initially read from disk, but then held in the kernel's page cache until read again or evicted to making room for more recently read records. The MongoDB server is run on a VM with 4 GB of local RAM, and in the FluidMem case 4 GB of off-node memory is added using memory hotplug. The FluidMem external RAM store used is RAMCloud.

\begin{figure*}[!htb]
	\centering
	\begin{subfigure}[t]{0.4\textwidth}
		\includegraphics[scale=0.45]{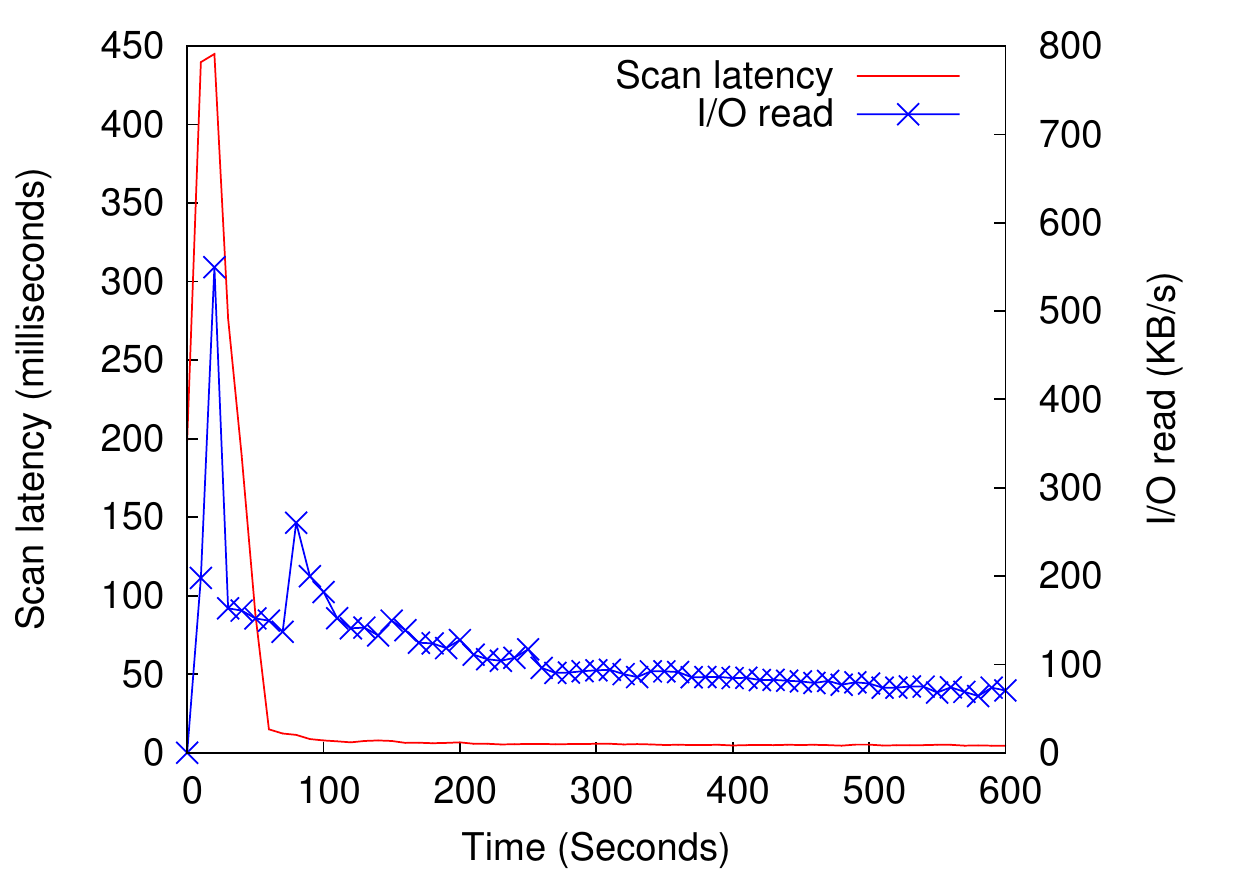} 
		\caption{Scan latency with 4GB FluidMem}
		\label{fig:fluidmem_scan_io.pdf}
	\end{subfigure}  	
	\begin{subfigure}[t]{0.4\textwidth}
		\includegraphics[scale=0.45]{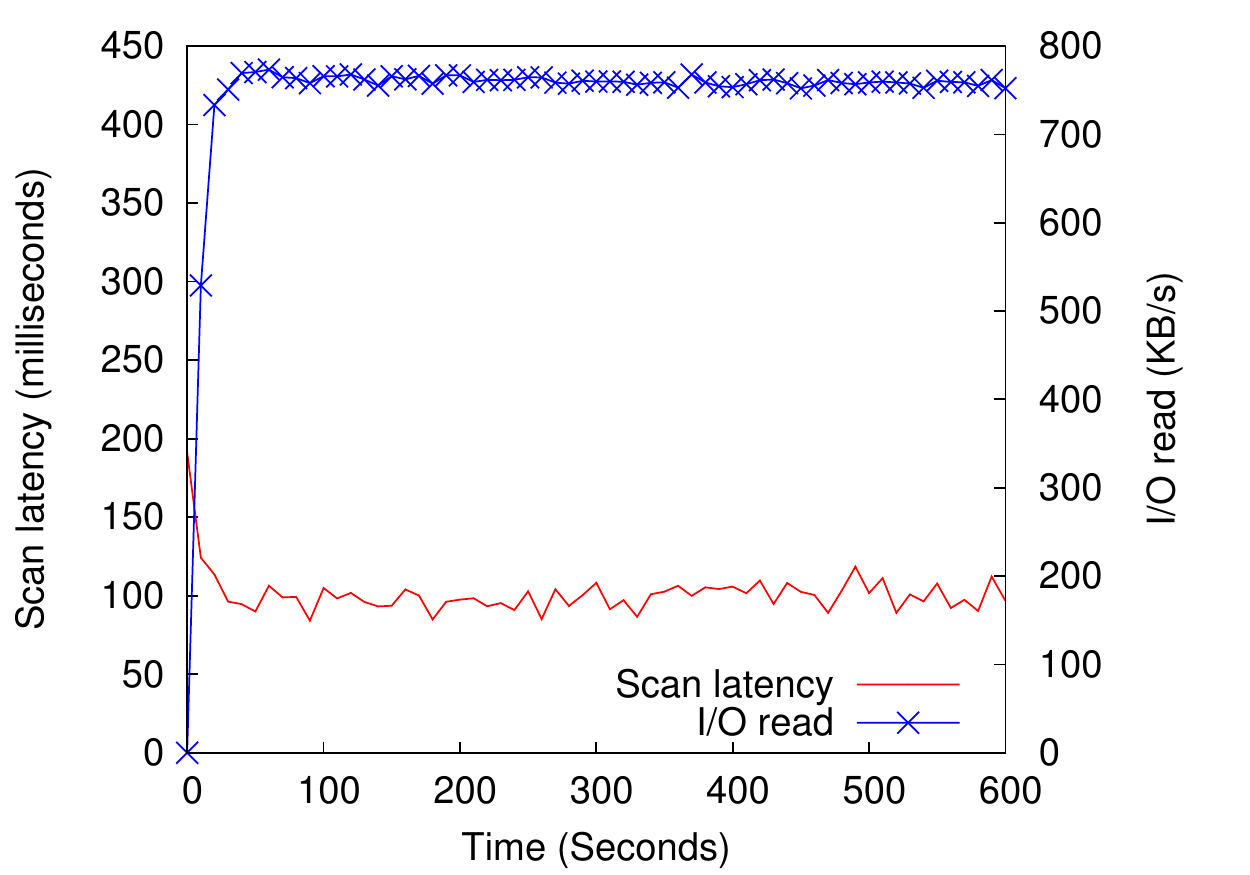} 
		\caption{Scan latency without FluidMem}
		\label{fig:swap_scan_io.pdf}
	\end{subfigure} \\
	\begin{subfigure}[t]{0.4\textwidth}
		\includegraphics[scale=0.45]{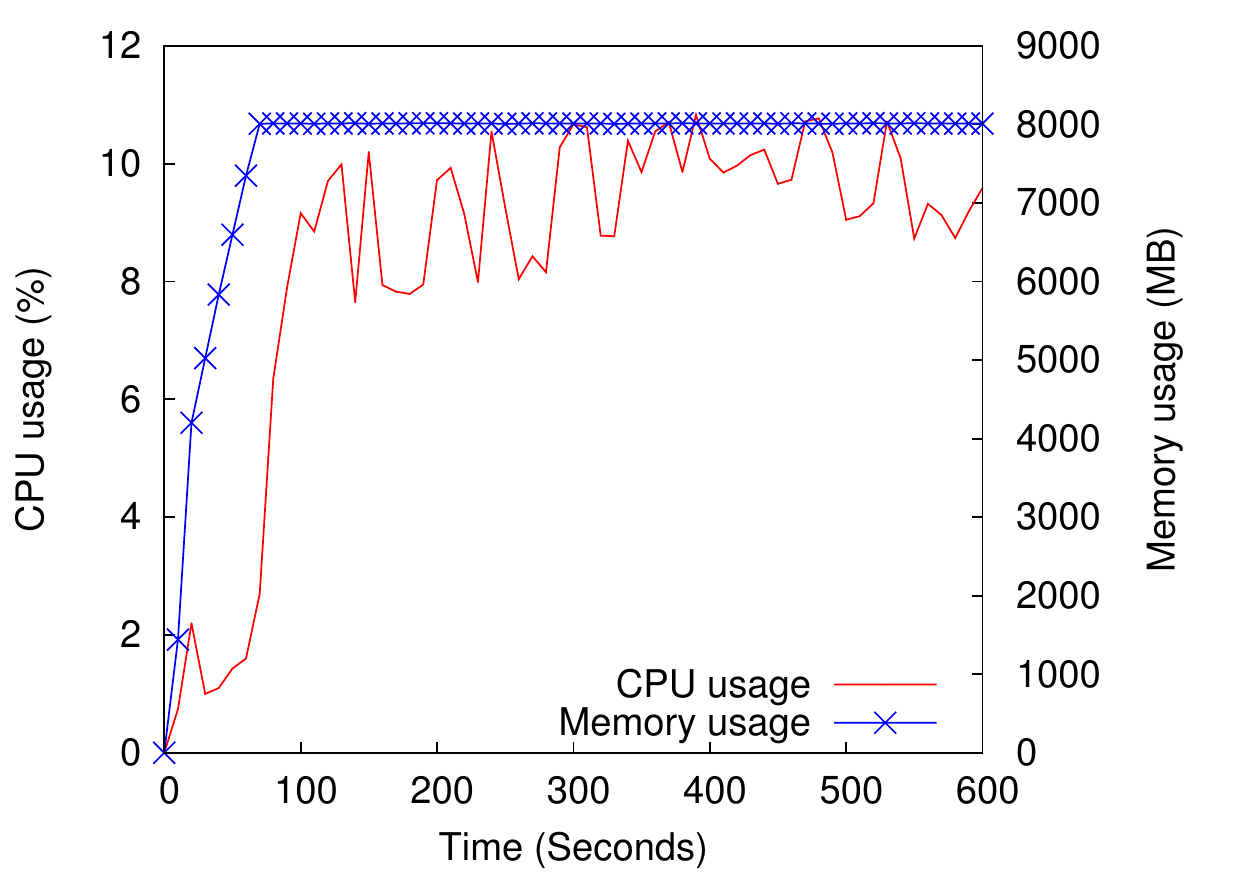} 
		\caption{CPU and memory usage with 4GB FluidMem}
		\label{fig:fluidmem_cpu_mem.pdf}
	\end{subfigure} 
	\begin{subfigure}[t]{0.4\textwidth}
		\includegraphics[scale=0.45]{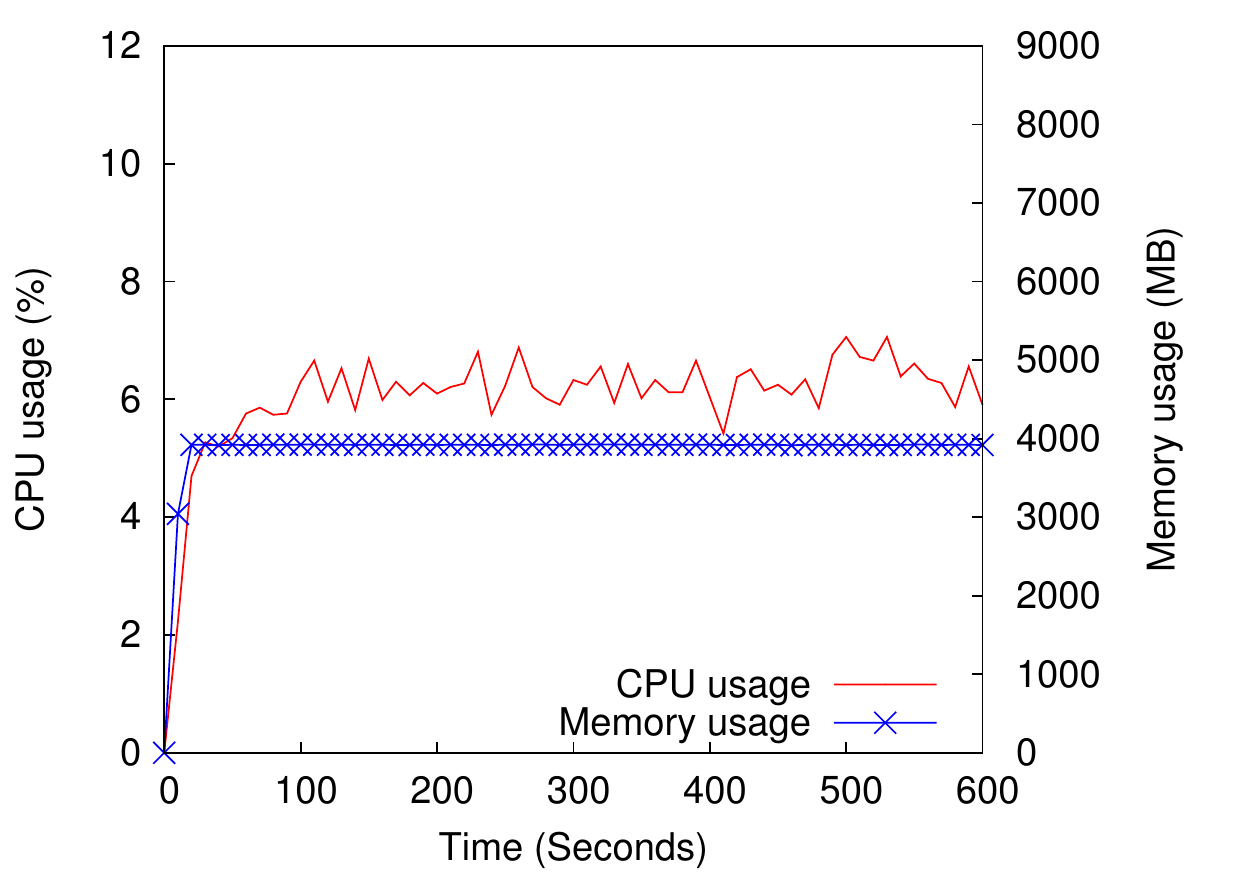} 
		\caption{CPU and memory usage without FluidMem}
		\label{fig:swap_cpu_mem.pdf}
	\end{subfigure}  
	
	\caption{MongoDB experiment results with and without FluidMem.}
	\label{fig:mongodb_result}
\end{figure*}

\begin{figure}[!htb]
	\centering
	\includegraphics[scale=0.42]{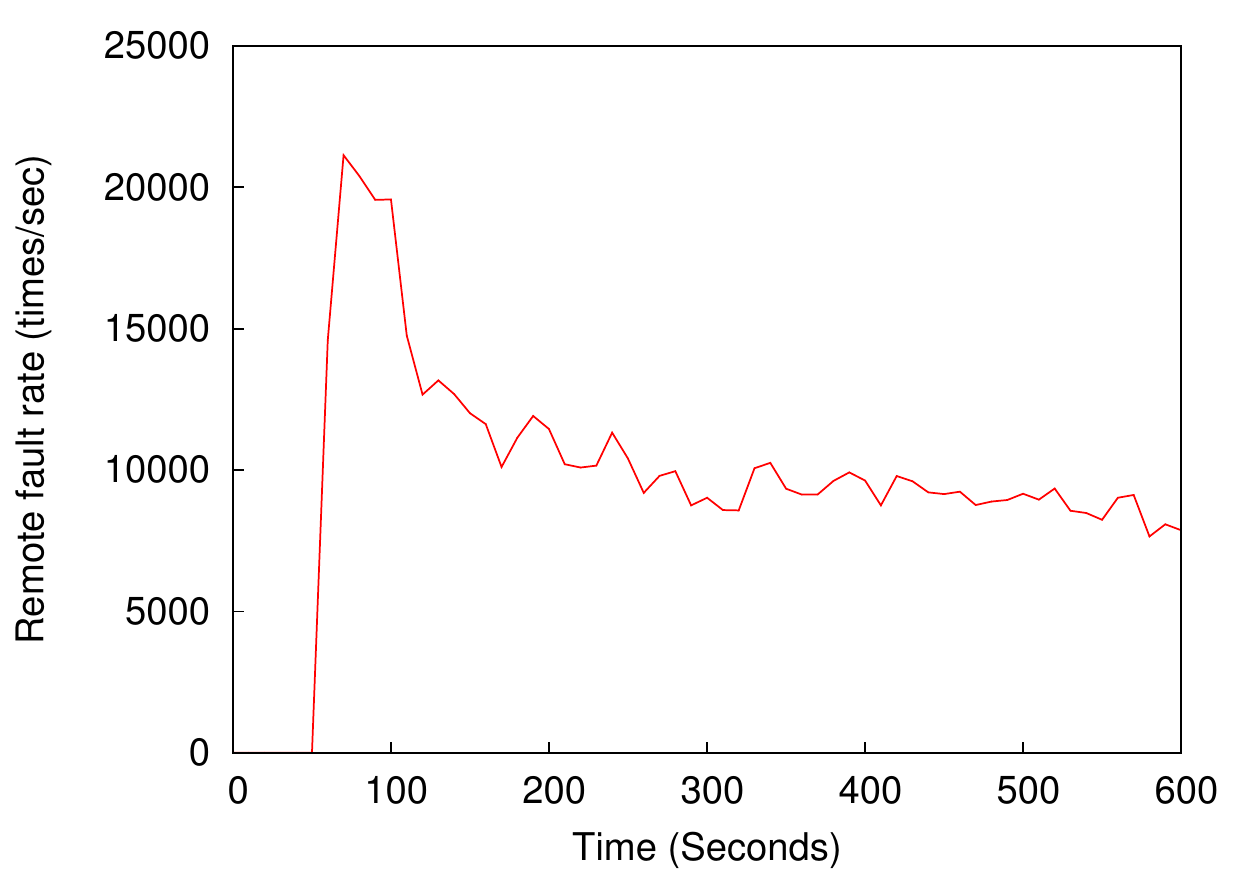}
	\caption{Remote fault rate of FluidMem in MongoDB experiment.}
	\label{fig:remote_fault_rate}
\end{figure}

Figure~\ref{fig:mongodb_result} shows the results of our MongoDB experiments with and without FluidMem.  In Figure ~\ref{fig:fluidmem_cpu_mem.pdf}, we see that FluidMem reaches a steady state using all 8 GB of memory, including remote memory for caching records. This is in contrast to the case without FluidMem where the kernel is only limited to 4 GB, as seen in Figure ~\ref{fig:swap_cpu_mem.pdf}.  With a smaller cache, more of the records need to be pulled from disk, resulting in much higher disk I/O. The effect on MongoDB operation is observed in Figures~~\ref{fig:fluidmem_scan_io.pdf} and ~\ref{fig:swap_scan_io.pdf}. With FluidMem, a greater number of the records can be retrieved from the in-memory cache, significantly reducing the amount of disk I/O (Figure~~\ref{fig:fluidmem_scan_io.pdf}).  However, some records need to be retrieved from FluidMem.  Figure~\ref{fig:remote_fault_rate} plots the remote page fault rate with FluidMem.  Even though remote page faults are incurred, the overall effect on average  latency is substantially reduced with FluidMem.  The average latency (for a 10KB record) drops from 100 ms without FluidMem to 5 ms with FluidMem for each 100 KB record.

\subsubsection{Genome assembly}

\label{sec:mongodb}

FluidMem offers a great benefit to some applications that would otherwise not run to completion given insufficient local memory.  By providing such applications with access to remote memory, FluidMem enables them to complete their execution without modification.  

In genomics, a large amount of memory is needed to accomplish DNA sequencing.
For a complete human genome, the construction of the de Bruijn graph using the memory-efficient SOAPdenovo assembler~\cite{Luo2012}, even after error reducing preprocessing steps took 140GB of RAM~\cite{Li2010}. Other assemblers would take terabytes of memory for a human-sized genome~\cite{Schatz2010}. Since the analysis of a de Bruijn graph is not easily parallelizable~\cite{Schatz2010} and the traversal does not exhibit significant spatial locality, the approach used by most assemblers is to construct and hold the entire graph in memory. Thus the maximum genome size that can be sequenced is limited by the amount of RAM on a single node.

Using FluidMem, two genome sequencing applications SOAPdenovo~\cite{Luo2012} and Velvet~\cite{zerbino:velvet:2008} are able to run to completion assembling the Human chromosome 14 from the GAGE dataset~\cite{Salzberg2012} on a VM with 72GB of memory (60GB local, 12GB FluidMem). Our nodes only have 64GB of DRAM each, so the only way we were able to complete the assembly without FluidMem was to combine the DIMMs from two machines for a node with 128GB of local DRAM.

\begin{table*}[th]
	\scriptsize
	\centering
	\begin{tabular}{|l|c|c|c|c|c|p{1.5in}|}
		\hline
		\multicolumn{1}{|c|}{}                                                             & \multicolumn{1}{l|}{\begin{tabular}[c]{@{}l@{}}No code \\ modification\end{tabular}} & OS support & \multicolumn{1}{l|}{\begin{tabular}[c]{@{}l@{}}Heterogenous \\ infrastructure\end{tabular}} & Isolation & \multicolumn{1}{l|}{\begin{tabular}[c]{@{}l@{}}Cloud \\ management\end{tabular}} &          \\ \hline
		\begin{tabular}[c]{@{}l@{}}Expand virtual memory \\ by modifying apps\end{tabular} & X                                                                                    & X          & X                                                                                           & X         &  \checkmark                                                                                & DSM~\cite{fleisch:sosp:1989,fleisch:spe:1994,souto:atc:1997,Bennett:1990:MDS}, PGAS~\cite{nelson_latency-tolerant_2015,Chamberlain:2007:PPC,Charles:2005:XOA}, key-value stores~\cite{dragojevic_farm:_2014,ousterhout:transactions:2015}, single system image (SSI)~\cite{www:kerrighed}, runtime framework~\cite{midorikawa:cluster:2008}  \\ \hline
		Cluster memory                                                                     & \checkmark                                                                                    & X          & X                                                                                           & X         & X                                                                                & never used in cloud data centers~\cite{feeley:sosp:1995,Chapman:2009:VVS} or proposing new ISA~\cite{Ma:2014:DVM}           \\ \hline
		\begin{tabular}[c]{@{}l@{}}Kernel-level \\ network swapping\end{tabular}           & \checkmark                                                                                    & \checkmark          & X                                                                                           & \checkmark         & X                                                                                & requiring a specific block device set up for each system~\cite{Flouris:1999:NRU,liang_swapping_2005,hines:vtdc:2007,newhall:euro-par:2003}           \\ \hline
		\begin{tabular}[c]{@{}l@{}}Rack-scale \\ memory disaggregation\end{tabular}        & \checkmark                                                                                    & \checkmark          & X                                                                                           & \checkmark         & \checkmark                                                                                & prototypes simulate hardware to manage the distributed storage of memory pages~\cite{han_network_2013,Gao:2016:OSDI,Lim:2009:DME:1555815.1555789,Lim:2012:SID:2192603.2192683}. RDMA support, but specialized distributed storage~\cite{gu:nsdi:2017},          \\ \hline
		FluidMem                                                                           & \checkmark                                                                                    & \checkmark          & \checkmark                                                                                           & \checkmark         & \checkmark                                                                                &            \\ \hline
	\end{tabular}
	\caption{Remote memory approaches and their limitations in terms of providing Memory as a Service (X = violation, checkmark = compliance).}
	\label{table:remotemem}
\end{table*}

\subsubsection{Graph500 and Spark}

Similar to the task in genome assembly of identifying Eulerian cycles of de Bruijn graphs, completing a breadth-first search (BFS) traversal is generally a memory-bound task due to irregular memory accesses~\cite{Buluc:2011}. For this reason we tested the sequential reference implementation of the Graph500 benchmark~\cite{murphy:2010}, which is a BFS graph traversal, on FluidMem.  Like the other applications, Graph500 ran successfully on FluidMem and was able to access remote memory without modification. Similarly, Apache Spark was deployed on our FluidMem infrastructure and successfully accessed remote memory without modification while completing execution.

\section{Related Work}
\label{sec:relates}

Table~\ref{table:remotemem} summarizes many of the memory disaggregation and remote memory approaches mentioned earlier. We have categorized these techniques by some of their most prominent characteristics.  The columns list each of the five requirements for Memory as a Service identified earlier.  
We first observe that many remote memory techniques, including DSM, PGAS, kev-value stores, SSI, etc., are characterized by some type of intervention on the part of the developer, requiring refactoring of their code to incorporate remote memory access.  Cluster-based memory approaches avoid this, but violate many of the other tenets of Memory as a Service.  Kernel-level network swap-based approaches hide the remote memory behind the block interface, thereby limiting the physical memory available to the VM.  We also believe that in order to flexibly support heterogeneous back ends that they will eventually have to move into user-space.  Rack-scale memory disaggregation approaches are primarily prototypes that do not currently support the heterogeneous complexity of today's datacenters.

\section{Conclusion}
\label{sec:concl}



In this paper, we have presented FluidMem, an implementation of Memory as a Service in the datacenter.  FluidMem leverages both the userfault handler code as well as the hotplug memory capability in the Linux kernel to provide transparent remote memory to standard applications such as MongoDB, genome sequencing, Graph500 and Spark.  Such applications may use FluidMem without modification.  MongoDB exhibits one of the benefits of FluidMem, namely that an application can improve its optimizations when the added capacity of the remote physical memory is exposed to the application.  The genome sequencing applications also benefitted from exposure to the additional remote memory by running to completion, since they would otherwise fail due to lack of local memory.

For future work, we hope to extend FluidMem to incorporate NUMA locality concepts and scale up the system to a larger deployment.
\section*{Acknowledgments}
\label{sec:acks}

This material is based upon work supported by the National Science
Foundation under Grant No. 1337399. Additionally, we would like the thank
William Mortl, Kannan Subramanian, and Daniel Zurawski for their work on
this project.

\bibliographystyle{acm}
\bibliography{./fluidmem.bib}


\end{document}